\documentclass[aps,prl,twocolumn,groupedaddress]{revtex4}
\usepackage{amsmath}
\usepackage{graphicx}
\DeclareGraphicsExtensions{.eps}
\newcommand{\ket}[2]{\mbox{$|#1\rangle_{#2}$}}
\newcommand{\bra}[2]{\mbox{$_{#2}\langle #1|$}}
\newcommand{\qgcnot}[2]{\ensuremath{\mbox{\textsf{CNot}}_{[#1]#2}}}
\newcommand{\qh}[1]{\ensuremath{\mbox{\textsf{H}}_{#1}}}
\newcommand{\qgcp}[2]{\ensuremath{\mbox{\textsf{CP}}_{#1}\left(#2 \right)}}

\begin{document}


\title{Suppressing decoherence of quantum algorithms by jump codes}


\author{O.~Kern}
\author{G.~Alber}
\affiliation{Institut f\"ur Angewandte Physik, Technische Universit\"at Darmstadt, D-64289 Darmstadt, Germany}


\date{\today}

\begin{abstract}
The stabilizing properties of one-error correcting jump codes are explored under realistic non-ideal conditions.
For this purpose the quantum algorithm of the tent-map is decomposed into a universal set of Hamiltonian quantum gates which ensure perfect correction of spontaneous decay processes under ideal circumstances even if they occur during a gate
operation.
An entanglement gate is presented which is capable of entangling any two logical qubits of different one-error correcting code spaces.
With the help of this gate simultaneous spontaneous decay processes affecting physical qubits of different code spaces can be corrected and decoherence can be suppressed significantly.
\end{abstract}


\maketitle

\section{Introduction\label{Introduction}}
Overcoming decoherence originating from  uncontrolled couplings between a quantum system and its environment
is one of the major challenges in the realization of quantum computers.
Numerous error correcting methods have been designed recently which are able to achieve this goal under ideal circumstances \cite{nielsenchuang}.
In particular, these ideal conditions require that
each error has to be corrected immediately and the correction itself has to be performed instantaneously.
In practice this requires
error detection, syndrome determination, and recovery operation to be executed on a time scale which is vanishingly small
compared to the intrinsic time scale of the quantum algorithm which is to be stabilized.
Typically it is difficult to fulfil these ideal conditions  and so the natural question arises
how do error correcting stabilization methods affect quantum algorithms under non-ideal conditions.

In this paper we explore this problem for the recently developed one-error correcting jump codes which have been designed for
correcting spontaneous decay processes of qubits \cite{alber1}. 
Jump codes are particularly useful in cases in which not only error times but also error positions are known.
They are based on an active error correcting quantum code \cite{shor1,knill,ekma,knlavi}
which is embedded in a decoherence free subspace \cite{zara,dugo,lidar}
in such a way that all errors taking place between successive spontaneous decay processes are corrected passively.
Thus, jump codes require a small number of recovery operations only and, in addition, 
within the family of all such embedded quantum codes
their redundancy
is minimal \cite{alber4}.
However, if one was able to control complex many-body Hamiltonians dynamically,
it would be possible to correct spontaneous decay processes
with quantum codes of even smaller redundancy
which involve one redundant qubit only 
\cite{wiseman}. 
Another proposal involving error correction of spontaneous decay processes with one redundant qubit was
explored in Ref. \cite{khod2}. But
this suggestion is erroneous as will be discussed later.
Thus, as long as it is still difficult to control complicated many-body Hamiltonians 
jump codes 
offer attractive perspectives for the correction of
spontaneous decay processes as their error correction involves
one- and two-qubit Hamiltonians only.

Applying jump codes to the stabilization of quantum algorithms one also ought to be able to correct spontaneous decay
processes which occur during the application of elementary quantum gates.
For this purpose one has to ensure that even during the application of a quantum gate the error correcting code space is not left at any time
\cite{alber5}.
This requirement can be fulfilled by realizing a universal set of
quantum gates with the help of Hamiltonians which leave an error correcting code space invariant. In addition, it is desirable
that these Hamiltonians are as simple as possible so that they can be realized in laboratory.
Recently, it was demonstrated that such Hamiltonian quantum gates can be constructed in a straightforward way
provided one restricts the encoding to appropriate subspaces of
one-error correcting jump codes \cite{khod1}. 
It is even possible to develop these Hamiltonian universal quantum gates
in such a way that
the logical qubits constituting these subspaces can be addressed individually, i.e. these logical subspaces can be equipped with a
natural tensor-product structure.

In the following we investigate the extent to which these latter one-error correcting jump codes
are capable of stabilizing quantum algorithms against spontaneous decay processes under non-ideal conditions.
As an example we consider
the recently proposed quantum algorithm of the tent-map \cite{shep144}.
Quantum maps of this kind provide interesting candidates of quantum algorithms which may be run on 
the first generations of few-qubit quantum computers \cite{shep129,shep135}.
Even if error times and error positions are known precisely and if the appropriate Hamiltonian quantum gates operate perfectly,
residual errors arise due to the finite duration of realistic recovery operations. In particular, any spontaneous decay process
occurring during a recovery operation cannot be corrected by an encoding within a one-error correcting jump code.
It is demonstrated that
the resulting decoherence can be suppressed significantly by using
a parallel encoding of the quantum registers of a quantum computer.
If the physical qubits of each quantum register constitute  a one-error correcting jump code,
for example,
all simultaneous or sequential spontaneous decay processes  can be corrected by such an encoding
provided they affect different quantum registers.
For this purpose we present
a universal entanglement gate which is capable of entangling any logical qubits of any two different
one-error correcting jump codes. 

This paper is organized as follows: In Sec.~II basic aspects of one-error correcting jump codes 
are summarized and a
universal set of quantum gates is discussed
whose one- and two-qubit Hamiltonians leave certain subspaces of these jump codes invariant.
In addition, a 
universal entanglement gate is presented which is capable of entangling any two logical qubits belonging to two different
error correcting code spaces.
In Sec. III the dynamics of the quantum algorithm of the tent-map are explored under the influence of realistic recovery operations
of finite duration.
It is demonstrated that decoherence can be suppressed significantly by a parallel encoding of the quantum registers
which also allows to correct simultaneous spontaneous decay processes affecting different error correcting code spaces.

\section{Spontaneous decay of qubits and one-error correcting jump-codes}
In this section basic aspects of the recently introduced one-error correcting jump codes are summarized. 
In particular, the recently proposed logical subspaces \cite{khod1} are discussed which can be equipped
with a tensor-product structure with the help of universal Hamiltonian quantum gates leaving these subspaces
invariant. A novel universal entanglement gate is presented
which is capable of entangling arbitrary logical qubits of different logical subspaces.
Decomposing quantum algorithms with the help of this quantum gate one can correct simultaneous
spontaneous decay processes provided they affect physical qubits of different error correcting code spaces.

Let us consider a typical quantum optical model of a quantum information processor
consisting of $n_q$ two level quantum systems (qubits) which can decay spontaneously
by emission of photons. If the distance between the qubits is much larger than
the wave length of the spontaneously emitted photons the resulting spontaneous decay
processes are statistically independent.
In the Born- and Markov-approximation
the dynamics of such a quantum system
can be described by a master equation in Lindblad form \cite{carmichael1}, i.e.
\begin{eqnarray}
\dot{\rho}_t &=& -\frac{i}{\hbar}[H_{sys}(t),\rho_t] +
\sum_{\alpha=0}^{n_q-1}(
[L_{\alpha},\rho_t L_{\alpha}^{\dagger}] +
[L_{\alpha}\rho_t, L_{\alpha}^{\dagger}]
).\nonumber\\
\label{Lindblad}
\end{eqnarray}
Thereby, $\rho_t$ denotes the reduced density operator of the $n_q$-qubit quantum system
at time $t$ and the Hamiltonian $H_{sys}(t)$ is assumed to describe the ideal dynamics of the
qubits due to a particular quantum algorithm.
The Lindblad operator 
\begin{equation}
L_\alpha=\sqrt{\kappa_\alpha}   \ket{0}{\alpha} \! \,_\alpha\bra{1}{} \otimes \openone_{\beta\neq \alpha},
\end{equation}
describes the spontaneous decay of qubit $\alpha$ 
from its (unstable) excited state $\ket{1}{\alpha}$ to its (stable) ground state 
$\ket{0}{\alpha}$ with the spontaneous decay rate $\kappa_{\alpha}$. 

\begin{figure}
\centering
\includegraphics[width=0.3\textwidth]{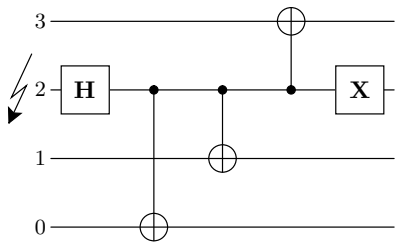}%
\caption{Graphical representation of the recovery operation $R_2$ which restores the quantum state of three logical qubits within a $(4,3,1)_2$-jump code after a spontaneous decay process of qubit $\alpha =2$.}
\label{fig::rec}
\end{figure}

For the important special case of equal decay rates, i.e. $\kappa_{\alpha} \equiv \kappa$ for $\alpha = 0,...,n_q -1$,
such a quantum information processor can be protected against spontaneous decay processes with the help
of the recently developed one-error detected-jump-error-correcting quantum codes or jump codes (JCs) \cite{alber1,alber2}.
These quantum codes correct all errors originating from the Lindblad operators of Eq.(\ref{Lindblad})
occurring between successive spontaneous decay processes passively with the help of appropriately
constructed decoherence free subspaces (DFSs) \cite{zara,dugo,lidar}.
In addition, provided error positions are known these quantum codes are
capable of correcting any single spontaneous emission event actively by an 
appropriately constructed active quantum code which is embedded within a DFS.
The orthonormal logical basis states (code words) of these quantum codes are 
constructed from all possible complementary pairings of the $n_q$-qubit states (with $n_q$ being even) which involve precisely
$n_q/2$ excited qubits.
In the case of $n_q =4$ the orthonormal basis states of the $(4,3,1)_2$ code, for example, are given by
\begin{subequations}
\begin{align}
\ket{c_0}{L}&=\frac{1}{\sqrt{2}} \bigl(\ket{0011}{}+\ket{1100}{}\bigr),\label{eq::431a}\\
\ket{c_1}{L}&=\frac{1}{\sqrt{2}} \bigl(\ket{0101}{}+\ket{1010}{}\bigr),\\
\ket{c_2}{L}&=\frac{1}{\sqrt{2}} \bigl(\ket{0110}{}+\ket{1001}{}\bigr)\label{eq::431c}\ .
\end{align}
\end{subequations}
Analogously, one may construct one-error correcting $(n_q,$ $\text{dim}_{n_q},1)_{n_q/2}$-codes
for any even number of physical qubits $n_q$ with the
dimension of the logical Hilbert space
\begin{equation}
\text{dim}_{n_q} = \frac{1}{2} \binom{n_q}{n_q/2}.
\end{equation}
This particular family of quantum  codes has the interesting property
that for any even number of physical qubits $n_q$ the redundancy of the associated JC is as small as possible
provided one aims at correcting errors between successive spontaneous emission events passively \cite{alber4}.
Furthermore, if a spontaneous decay process of qubit $\alpha$ has been detected the resulting error can be
corrected by applying the unitary recovery operation
\begin{eqnarray}\label{eq::rec}
R_\alpha= X_{\alpha}  \biggl( \prod_{\beta \neq \alpha} \qgcnot{\alpha}{\beta} \biggr) \qh{\alpha},
\end{eqnarray}
as illustrated in figure \ref{fig::rec}.
Thereby, $X_{\alpha}$, $\qgcnot{\alpha}{\beta}$, and $\qh{\alpha}$ denote a 
not-gate acting on qubit $\alpha$,
a controlled not-gate acting on qubit $\beta$
with control qubit $\alpha$, and a Hadamard-gate acting on qubit $\alpha$.

In order to use a one-error correcting $(n_q, \text{dim}_{n_q},$ $1)_{n_q/2}$-code
for the perfect stabilization of arbitrary quantum algorithms 
one has to develop appropriate sets of universal quantum gates which guarantee that
the code space is not left even during the application of these gates \cite{alber5}.
Otherwise any spontaneous emission process taking place during the
application of an elementary quantum  gate can no longer be corrected.
This requirement can be achieved with the help of Hamiltonian quantum gates provided
the relevant Hamiltonians leave the relevant code space invariant.
So far, for general one-error correcting jump codes the construction of such universal quantum gates has been possible
in special cases only \cite{alber5}. However, restricting oneself to appropriate subspaces of these one-error correcting 
jump codes 
it is possible to construct such universal quantum gates in a straightforward way at the prize of increasing redundancy.
Examples of such subspaces have been proposed recently by 
Khodjasteh and Lidar \cite{khod1}. In the case of $n_q$ physical qubits
such an appropriate subspace of the one-error correcting 
$(n_q, \text{dim}_{n_q}, 1)_{n_q/2}$-code
is spanned by the orthonormal logical states
\begin{equation}\label{KL}
\begin{split}
\ket{00\dots 0}{L}&=\frac{1}{\sqrt{2}}\bigl(\ket{01}{}\otimes\ket{01}{}\otimes\dots\otimes\ket{01}{}\otimes\ket{01}{}+\\
&\mathrel{\hphantom{=}}\hphantom{\frac{1}{\sqrt{2}}\bigl(}\ket{10}{}\otimes\ket{10}{}\otimes\dots\otimes\ket{10}{}\otimes\ket{10}{}\bigr),\\
\ket{00\dots1}{L}&=\frac{1}{\sqrt{2}}\bigl(\ket{01}{}\otimes\ket{01}{}\otimes \dots\otimes\ket{10}{}\otimes \ket{01}{}+\\
&\mathrel{\hphantom{=}}\hphantom{\frac{1}{\sqrt{2}}\bigl(}\ket{10}{}\otimes\ket{10}{}\otimes\dots\otimes\ket{01}{}\otimes\ket{10}{}\bigr),\\
&\vdots\\
\ket{11\dots 1}{L}&=\frac{1}{\sqrt{2}}\bigl(\ket{10}{}\otimes\ket{10}{}\otimes\dots\otimes\ket{10}{}\otimes\ket{01}{}+\\
\underbrace{\hphantom{\ket{11\dots 1}{L}}}_{\substack{\text{$n_{\!L}$ logical}\\\text{qubits}}}
&\mathrel{\hphantom{=}}\mathrel{\hphantom{ \frac{1}{\sqrt{2}}\bigl( }}
\underbrace{\ket{01}{}\otimes\ket{01}{}\otimes\dots\otimes\ket{01}{}}_\text{$2n_{\!L}$ physical qubits} 
\otimes \ket{10}{} \bigr)
\end{split}
\end{equation}
for example. In this case
$n_q$ physical qubits are required for the encoding of $n_{\!L} = (n_q-2)/2$ logical qubits. 
Besides the complementary pairing of the $n_q$-qubit states 
Eq.(\ref{KL}) also involves a complementary pairing of adjacent physical qubits. Thereby the rightmost two physical qubits are used
for distinguishing the complementary states of each pair of $n_q$-qubit states.
In the code space spanned by
the orthonormal logical states  of Eq.(\ref{KL})  any spontaneous decay process of any of the $n_q$ physical qubits
can be corrected provided error position and error time are known.

\subsection{Universal sets of quantum gates within one-error correcting code spaces}
How can we develop a Hamiltonian set of universal quantum gates which do not leave the code space of Eq.(\ref{KL})
at any time? This can be achieved with the help of the elementary Ising- and Heisenberg-type two-qubit Hamiltonians
$T_{\alpha \beta}=(X_{\alpha} \otimes X_{\beta} + Y_{\alpha} \otimes Y_{\beta})/2$ and $ZZ_{\alpha \beta} \equiv Z_{\alpha} \otimes Z_{\beta}$,
for example,
which act on the physical qubits $\alpha$ and $\beta$. (The operators $X, Z$, and $Y$ denote the three anti-commuting  Pauli
spin operators of the appropriate qubits with $XY = i Z$ etc.)
Provided one is capable of controlling these 
two-qubit Pauli
Hamiltonians 
it is straightforward to construct a universal set of logical single and two-qubit Hamiltonians
$\overline{X}_i,\overline{Z}_i$, and $\overline{ZZ}_{ij}$
which act on the logical qubits $i$ and $j$ 
similarly as the corresponding Hamiltonians $T_{\alpha \beta}, ZZ_{\alpha \beta}$
and which leave the code space of $n_{\!L}$ logical qubits
invariant \cite{khod1}, i.e.
\begin{subequations}
\begin{align}
\overline{X}_i&=T_{2i+3,2i+2},\label{eq::hx}\\
\overline{Z}_i&=ZZ_{2i+3,1},\label{eq::hz}\\
\overline{ZZ}_{ij}&=Z_{2i+3}\otimes Z_{2j+3}.\label{gates1}
\end{align}
\end{subequations}
With the help of these Hamiltonians any two logical qubits $i$ and $j$ can be addressed individually thus
providing a tensor product structure in the $2^{n_{\!L}}$-dimensional Hilbert space spanned by the basis state of
Eq.(\ref{KL}).
In Eq.(\ref{KL}) it is assumed that the logical (physical) qubits
are numbered from right to left from $i,j=0$ ($\alpha,\beta =0$) up to the maximum value of $i,j= n_{\!L} -1$ ($\alpha,\beta = n_q -1$).
The names of these
Hamiltonians indicate how they act on the logical states. Thus, the Hamiltonian
$\overline{ZZ}_{ij}$, for example, acts on the logical qubits $i$ and $j$ in the same way
as the Hamiltonian $ZZ_{\alpha\beta}$ acts on the physical qubits $\alpha$ and $\beta$.

Quantum gates built with the help of these Hamiltonians always stay within the code space spanned by the logical
basis of Eq.(\ref{KL}) for any even number of physical qubits.
The construction of universal logical single and two-qubit gates, such as Hadamard- and phase gates, based on these Hamiltonians
is exemplified in Figs. 2 and 3.
\begin{figure}
\centering
 \includegraphics[scale=0.85]{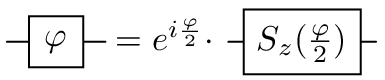}\\
 \includegraphics[scale=0.85]{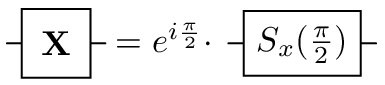}\\
 \includegraphics[scale=0.85]{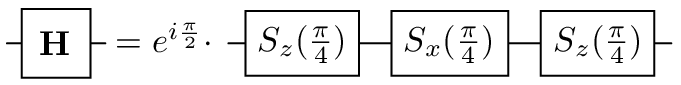}
 \caption{Single qubit gates constructed with $X$- and $Z$-type Hamiltonians with $S_{x}(\tau)=\exp ( -i X \tau)$ etc..
These Hamiltonians may represent either the Pauli operators $X$ and $Z$ or 
the logical Hamiltonians $\overline{X}$ and $\overline{Z}$ of Eqs.(\ref{eq::hx}) and (\ref{eq::hz}).
From top to bottom: Phase-gate ($\ket{0}{}\!\bra{0}{}+e^{i \varphi} \ket{1}{}\!\bra{1}{}$),
not-gate ($\ket{0}{}\!\bra{1}{}+ \ket{1}{}\!\bra{0}{}$) and Hadamard-gate
$( (\ket{0}{}+\ket{1}{}\!)\bra{0}{} + (\ket{0}{}-\ket{1}{}\!)\bra{1}{} )/\sqrt{2} $.}
\label{fig::sqgates}
\end{figure}
\begin{figure*}
 \hfill\includegraphics[scale=0.82]{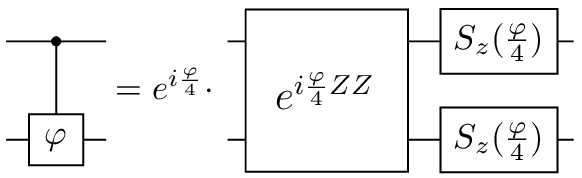}%
 \hfill\includegraphics[scale=0.82]{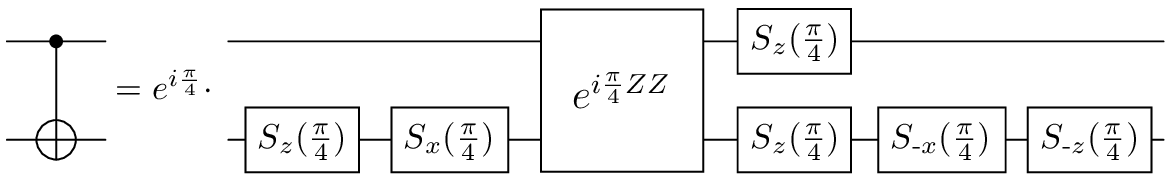}\hspace*{\fill}%
 \caption{Two qubit gates 
constructed with $X$-, $Z$- and $ZZ$-type Hamiltonians with $S_{x}(\tau)=\exp ( -i X \tau)$ etc..
These Hamiltonians may represent either the Pauli operators $X$, $Z$, and $ZZ$ or the logical Hamiltonians $\overline{X}$, $\overline{Z}$, and $\overline{ZZ}$ of Eqs. (\ref{eq::hx}), (\ref{eq::hz}), and (\ref{gates1}).
From left to right: Controlled phase-gate and controlled not-gate.}
\label{fig::tqgates}
\end{figure*}
Furthermore,
as the code space of Eq.(\ref{KL}) is a subspace of a $(n_q,{\rm dim}_{n_q},1)_{n_q/2}$ one-error correcting jump code
any detected spontaneous decay of any physical qubit $\alpha$  can be corrected by the recovery operation of Eq.(\ref{eq::rec}).
On this occasion we want to point out that the recovery operation proposed in Ref. \cite{khod1} does not work properly
since it does not even restore the basis states of Eq.(\ref{KL}).
This mistake might have its origin in a misinterpretation of the jump operator
$\ket{0}{}\!\bra{1}{}$ which actually transforms the physical state $\ket{0}{}$ into the zero-vector $0$ (which is not a physical state)
and not into $\ket{0}{}$ again.
The same misinterpretation appears to lead to the construction of a wrong code
space in a subsequent paper by the same authors \cite{khod2}.

\subsection{A universal entanglement gate for different one-error correcting code spaces}
The one-error correcting quantum codes of Eq.(\ref{KL}) cannot correct
simultaneous spontaneous decay processes affecting different physical qubits.
However, this problem 
can be overcome at least partly
by combining one-error correcting quantum codes each of which involves
a relatively small number of physical qubits.
Physically this can be achieved by a local architecture of a quantum information processor, for example,
which is based on small (local) groups of physical qubits (elementary registers) each of which constitutes a one-error correcting
code space as described by Eq.(\ref{KL}).
As a result
any number of simultaneous decay processes
can be corrected in parallel provided these decays affect physical qubits of
different one-error correcting code spaces.
In order to be able to stabilize arbitrary quantum algorithms
in such a quantum information processor 
one has not only to be able to perform arbitrary unitary operations within each elementary quantum register but one also has to
be able 
to entangle any two logical qubits of any to different elementary quantum registers.
In the following we construct such a universal entanglement gate.

For this purpose let us consider first of all the simple case of the tensor product space of two logical qubits.
According to Eq.(\ref{KL}) the resulting four basis states are given by
\begin{equation}\label{eq::tens}
\begin{split}
\ket{0}{La}\otimes\ket{0}{Lb}&=\frac{1}{2}\bigl( \ket{01}{}\otimes\ket{01}{}+\ket{10}{}\otimes\ket{10}{}\bigr)_a \otimes\\
&\mathrel{\hphantom{=}}\hphantom{\frac{1}{2}}\bigl( \ket{01}{}\otimes\ket{01}{}+\ket{10}{}\otimes\ket{10}{}\bigr)_b,\\
\ket{0}{La}\otimes\ket{1}{Lb}&=\frac{1}{2}\bigl( \ket{01}{}\otimes\ket{01}{}+\ket{10}{}\otimes\ket{10}{}\bigr)_a \otimes\\
&\mathrel{\hphantom{=}}\hphantom{\frac{1}{2}}\bigl( \ket{10}{}\otimes\ket{01}{}+\ket{01}{}\otimes\ket{10}{}\bigr)_b,\\
\ket{1}{La}\otimes\ket{0}{Lb}&=\frac{1}{2}\bigl( \ket{10}{}\otimes\ket{01}{}+\ket{01}{}\otimes\ket{10}{}\bigr)_a \otimes\\ 
&\mathrel{\hphantom{=}}\hphantom{\frac{1}{2}}\bigl( \ket{01}{}\otimes\ket{01}{}+\ket{10}{}\otimes\ket{10}{}\bigr)_b,\\
\ket{1}{La}\otimes\ket{1}{Lb}&=\frac{1}{2}\bigl( \ket{10}{}\otimes\ket{01}{}+\ket{01}{}\otimes\ket{10}{}\bigr)_a \otimes\\
&\mathrel{\hphantom{=}}\hphantom{\frac{1}{2}}\bigl( \ket{10}{}\otimes\ket{01}{}+\ket{01}{}\otimes\ket{10}{}\bigr)_b
\end{split}
\end{equation}
with the subscripts $a$ and $b$ referring to two different quantum registers.
The Hamiltonian
\begin{equation}
H^{a,b}_\text{ent}={ZZ}_{4,0} + {ZZ}_{6,1} + ZZ_{5,2} + ZZ_{7,3}
\label{ent}
\end{equation}
has the following properties
\begin{eqnarray}
&H^{a,b}_\text{ent}&\ket{m}{La}\otimes \ket{n}{Lb} = 0,~~(m,n)\neq (1,1)\nonumber\\
&H^{a,b}_\text{ent}& \ket{u}{a,b}=
4\ket{u}{a,b},\nonumber\\
&H^{a,b}_\text{ent}& \ket{v}{a,b}=
-4\ket{v}{a,b}\nonumber\\
\end{eqnarray}
with
$\ket{u}{a,b} =(\ket{10}{}\otimes\ket{01}{})_a \otimes (\ket{10}{}\otimes\ket{01}{})_b  +
(\ket{01}{}\otimes\ket{10}{})_a \otimes (\ket{01}{}\otimes\ket{10}{})_b$,
$\ket{v}{a,b} =(\ket{10}{}\otimes\ket{01}{})_a \otimes (\ket{01}{}\otimes\ket{10}{})_b  +
(\ket{01}{}\otimes\ket{10}{})_a \otimes (\ket{10}{}\otimes\ket{01}{})_b$.
Thus, this Hamiltonian generates a controlled $\pi$-phase gate
\begin{equation}
\qgcp{}{\pi} \equiv \openone-2 \ket{11}{}\bra{11}{}=
\exp \Bigl( -i H^{a,b}_\text{ent} \frac{\pi}{4} \Bigr).
\label{ent2}
\end{equation}
Since the four $ZZ$ Hamiltonians appearing in Eq.(\ref{ent}) commute, they can also be applied to the physical qubits one after the other.
Note that the gate of Eq.(\ref{ent2}) has similarities with the entanglement gate presented in Ref. \cite{alber5} for entangling
the qubits of two one-error correcting $(4,3,1)_2$-code spaces.

\begin{figure}
\centering
\includegraphics[width=0.475\textwidth]{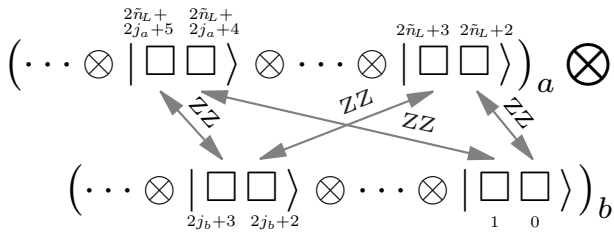}%
\caption{Entanglement gate which performs a \qgcp{}{\pi}-gate between a pair $(j_a,j_b)$ of logical qubits located in different jump codes $a$ and $b$ with $n_{\!L}$ and $\tilde{n}_L$ logical qubits.}
\label{fig::entgate}
\end{figure}
\begin{figure*}
\centering
\includegraphics[scale=0.82]{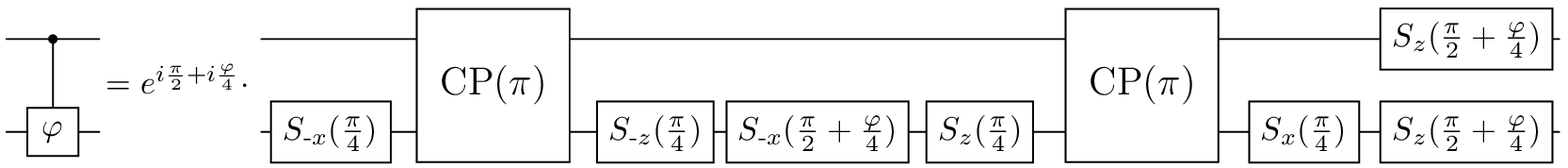}\\
\vspace{0.5cm}
\includegraphics[scale=0.82]{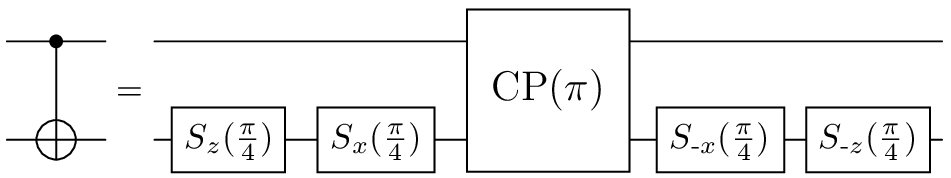}
\caption{The controlled $\pi$-phase gate (\qgcp{}{\pi}) can be used to construct any unitary quantum gate:
Construction of a controlled-phase-gate (top) and of a controlled-not-gate (bottom).}
\label{fig::cpgates}
\end{figure*}

The Hamiltonian entanglement gate of Eq.(\ref{ent2}) can be generalized to two jump codes of Eq.(\ref{KL}) of arbitrary sizes.
In particular, a controlled $\pi$-phase gate affecting the  logical qubits $j_a$ and $j_b$ of the elementary registers $a$ and $b$ which contain
$n_{\!L}$ and $\tilde{n}_L$ logical qubits is given by Eq.(\ref{ent2}) with the entangling Hamiltonian (compare with Fig.\ref{fig::entgate})
\begin{eqnarray}
H^{j_a,j_b}_\text{ent} &=& ZZ_{2\tilde{n}_L +2,0} + ZZ_{2j_a+2\tilde{n}_L+4,1} +\nonumber\\
&& ZZ_{2\tilde{n}_L +3,2j_b+2}+ZZ_{2j_a+2\tilde{n}_L+5,2j_b+3}.
\label{gate3}
\end{eqnarray}
Together with the universal quantum gates based on the Hamiltonians of Eqs.(\ref{eq::hx}),(\ref{eq::hz}),(\ref{gates1})
this controlled $\pi$-phase gate
constitutes a universal set of elementary quantum gates \cite{brylinski} with which any unitary transformation can be realized in a quantum information processor
with a local architecture.
Two examples of such unitary transformations and their corresponding decompositions in terms of these universal quantum gates
are depicted in Fig. \ref{fig::cpgates}.

There is one particularity concerning the recovery operation which must be applied to restore the code space after a
detected spontaneous decay taking place
during the application of the entanglement gate of Eq.(\ref{ent2}).
In general during the application of this entanglement gate a quantum state
leaves the tensor product space of the associated elementary quantum registers. But 
a perfect recovery 
by an appropriate gate sequence \eqref{eq::rec} is still possible.
This is due to the fact that
even during the application of an entanglement gate of the form of Eq.(\ref{gate3})
an arbitrary linear superposition of basis states of Eq.(\ref{KL}) always remains inside
the code space of a one-error correcting jump code.

%

\section{Stabilization of quantum maps under non-ideal conditions}
The jump codes discussed in the previous section 
provide perfect protection against spontaneous decays of qubits provided
error times and error positions are known precisely and appropriate recovery operations are applied immediately and instantaneously.
In this section we investigate the dynamics of quantum algorithms 
whose decoherence with respect to spontaneous decay
processes is stabilized by the one-error correcting jump codes discussed in Sec. II.
In particular, we are mainly interested in situations in which
a quantum algorithm is implemented ideally by a universal set of quantum gates which do not leave an error correcting
code space at any time. This way one is able to correct spontaneous decay processes even if they occur during the application of 
an elementary quantum gate.
However,
as a result of such an encoding
the time required for the application of a recovery operation is typically no longer
negligibly small in comparison with the intrinsic time evolution of a quantum algorithm so that the ideal conditions of error correction
are no longer fulfilled.

In order to explore the consequences of such non-ideal conditions let us consider the recently proposed quantum algorithm of the
iterated tent-map \cite{shep144} as an example. 
In each iteration of the  quantum tent-map an initial quantum state $\ket{\psi}{}$ is mapped onto the state
\begin{eqnarray} 
\ket{\psi'}{} &=& e^{-iT \hat{p}^2/2} e^{-ik\hat{V}(\hat{x})}\ket{\psi}{}
\end{eqnarray} 
with the tent-shaped force
\begin{eqnarray}
-V^{\prime}(x) = \left\{
\begin{array}{ll}
 (x -\frac{\pi}{2}),&(0\leq x<\pi)\\
 (\frac{3\pi}{2} - x),&(\pi \leq x< 2\pi).
\end{array}\right.
\end{eqnarray}
Thereby, $\hat{p}$ and $\hat{x}$ are dimensionless quantized action-angle variables.
Universal gate sequences involving Hadamard-, phase-, controlled phase-, and controlled not-gates
were developed recently \cite{shep144}.
In order to stabilize this quantum algorithm against spontaneous decay processes these
universal quantum gates and the recovery operations required have to be decomposed in terms of the elementary gates discussed
in the previous section which do
not leave the error correcting code space at any time
(compare with Figs.\ref{fig::sqgates}, \ref{fig::tqgates}, \ref{fig::cpgates}).

In the following we present numerical simulations for $t=30$ iterations of the quantum tent-map involving $n_{\!L}=6$ logical qubits. 
The characteristic parameters chosen are $k=1.7/T$ and $T=2\pi/2^{n_{\!L}}$.
As an initial quantum state we chose a coherent state centered around the mean value $(x,p)=(5.35,0)$ \cite{shep144}.
We compare the dynamics of the quantum algorithm which is stabilized by
the one-error correcting jump code of Eq.(\ref{KL}) involving $n_q=2 n_{\!L} +2 =14$ physical qubits with the corresponding
dynamics of $n_q=n_{\!L} = 6$ qubits in the absence of error correction for different spontaneous decay rates $\kappa$.
With this choice of parameters each iteration of
the quantum tent-map can be decomposed into $n_g=125$ elementary Hadamard-, phase-, controlled phase-, and controlled
not-gates acting on these $n_{\!L}$ logical qubits \cite{shep144}.
In our numerical simulations
each of these $n_g$ quantum gates is decomposed into a suitable sequence of universal logical quantum gates 
according to the gate libraries depicted in Figs. \ref{fig::sqgates} and \ref{fig::tqgates}. 
Thus, for each iteration of the quantum tent-map $437$ of these universal quantum gates are required
whose
Hamiltonians
have to be turned on for appropriate values
of the dimensionless parameter $\tau$. This latter parameter
may be viewed as a dimensionless measure for the duration of each of these elementary universal quantum gates.
Adding the $\tau$-values of the gates needed for one iteration of the quantum map
one obtains the value $\tau_{it} = 67.2\pi$
which implies
an average dimensionless time of magnitude $\tau_{it}/n_g = 0.54\pi$ for each of the 
$n_g$ quantum gates of Ref. \cite{shep144}.

The numerical results are obtained by simulating the master equation of Eq.(\ref{Lindblad}) with the quantum trajectory method \cite{dumpazo}. 
As soon as a spontaneous decay process takes place the relevant Hamiltonian, say $\hat{H}_0$,
implementing the actual logical quantum gate is stopped
immediately and the Hamiltonians are turned on which are required to perform the relevant recovery operation. 
Summing up the $\tau$-values of a recovery operation yields 
$\tau_\text{rec}= n_q\tau_\text{cnot} - \pi/2= 24 \pi$ (for $n_q = 14$) with the 
duration of a CNOT-operation being given by $\tau_\text{cnot} = 7\pi/4$ 
(compare with
Eq. \eqref{eq::rec} and Figs. \ref{fig::sqgates} and \ref{fig::tqgates}).
After completion of the recovery operation the quantum algorithm is started again with the previously stopped 
Hamiltonian $\hat{H}_0$. Of course, spontaneous decay processes occurring during a recovery operation cannot be corrected.
The resulting non-ideal performance of the quantum algorithm can be measured
by the fidelity
\begin{equation}
{f(t)}= {\rm Tr}(\rho_t \ket{\overline{\psi}(t)}{}\bra{\overline{\psi}(t)}{})
\end{equation} 
of the non-ideal quantum state $\rho_t$
with respect to
the corresponding ideal (pure) quantum state
$\ket{\overline{\psi}(t)}{}$.

\begin{figure}
\centering
\includegraphics[width=0.5\textwidth, trim=14 0 0 0]{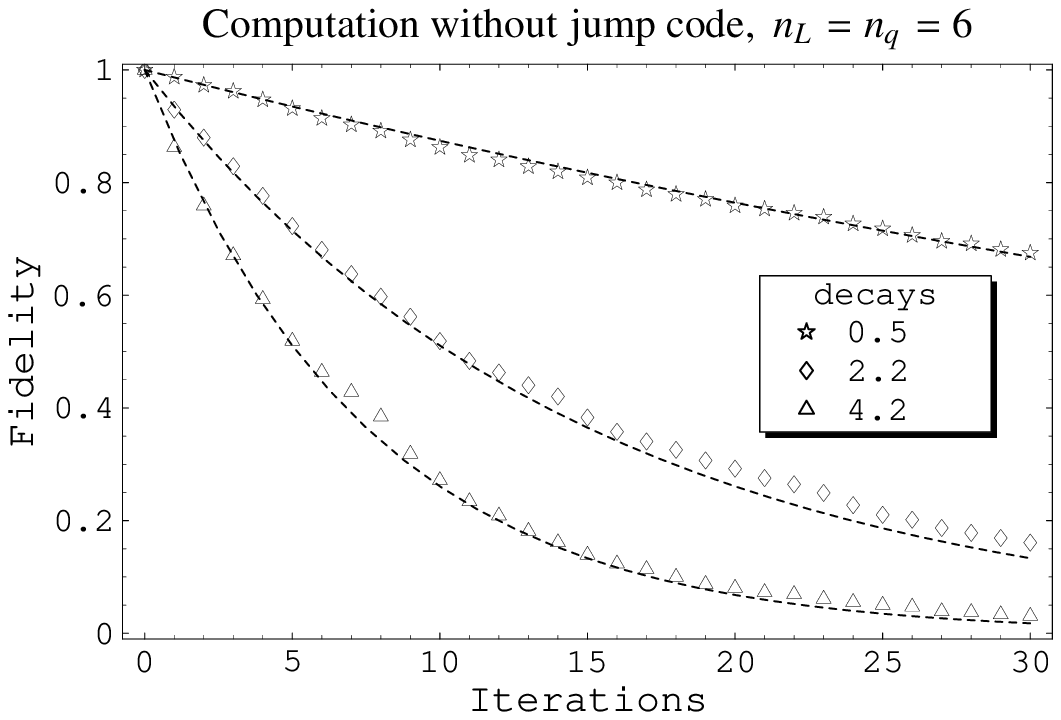}\\
\vspace{0.4cm}
\includegraphics[width=0.5\textwidth, trim=14 0 0 0]{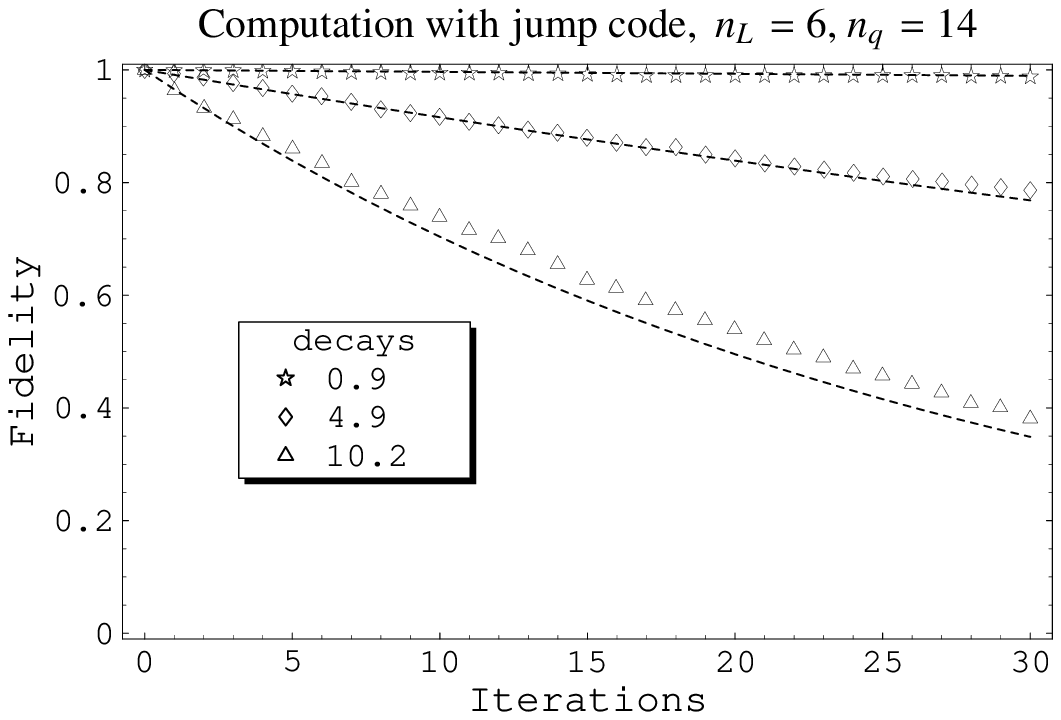}
\caption{Time evolution of the fidelity of the quantum tent-map for three different spontaneous decay rates:
${\kappa}= 2/(3\pi) \times 10^{-4}$ (stars),
${\kappa}=10/(3\pi)\times  10^{-4}$ (diamonds),
and  ${\kappa}=2/(3\pi) \times 10^{-3}$ (triangles).
The corresponding mean numbers of spontaneous decays processes after 30 iterations are depicted in the insets.
Upper diagram: no error-correction, $n_{\!L} = n_q =6$;
lower diagram: with error correction, $n_{\!L} =6, n_q =14$.
The  dashed lines indicate the corresponding approximate fidelities of Eqs. (\ref{approxwithout}) and (\ref{approxwith}).}
\label{fig::pjc}
\end{figure}

In Fig. \ref{fig::pjc} the time evolution of this fidelity is depicted for different values of the (dimensionless)
spontaneous decay rate $\kappa$ of 
the qubits without (upper diagram) and with (lower diagram)
error correction. The  fidelities presented were averaged over $10^3$ statistical realizations.
In the upper diagram of Fig. \ref{fig::pjc}
$\ket{\overline{\psi}(t)}{}$ denotes the ideal quantum state of a quantum computer consisting of $n_q=6$ physical qubits
and
$\rho_t$ is the corresponding quantum state in the presence of spontaneous emission.
In the lower diagram of Fig.\ref{fig::pjc}
$\ket{\overline{\psi}(t)}{}$ is the ideal quantum state of a quantum computer consisting of $n_q=14$ physical qubits in the absence
of spontaneous decay processes.
Analogously, $\rho_t$ is the quantum state of a quantum computer involving $n_q=14$ physical and
$n_{\!L} = (n_q - 2)/2 = 6$ logical qubits whose dynamics are stabilized against spontaneous decay processes by
an appropriate encoding with the help of the one-error correcting quantum code of Eq.(\ref{KL}) and the quantum gates of Figs. 
\ref{fig::sqgates} and \ref{fig::tqgates}.

It is apparent from the dashed curves in Fig. \ref{fig::pjc} that the numerically obtained fidelities 
are described approximately by the relations
\begin{equation}
f_\text{er}(t)=\exp \left( - \frac{n_q}{2} {\kappa}  \tau_\text{it}  t \right)
\label{approxwithout}
\end{equation} 
without error correction and by
\begin{equation}
f_\text{ec1}(t)=\exp \left( - \left( \frac{n_q}{2} {\kappa} \right)^2   \tau_\text{rec}  \tau_\text{it} t
\right)\equiv
e^{-R_\text{ec1} \tau_\text{it} t},
\label{approxwith}
\end{equation}
in the case of error correction.
Thereby, $t$ denotes the number of iterations of the quantum map and $R_\text{ec1}$ is the fidelity decay rate in the presence
of error correction.
The quantity $f_\text{er}(t)$ of
Eq.(\ref{approxwithout})
resembles the recently proposed formula of Ref. \cite{shep152} where
the elementary quantum gates were performed instantaneously. In our notation this corresponds to
the case $\tau_\text{it}=n_g$. According to Eq.(\ref{approxwithout})
the mean fidelity
is estimated by the probability that no spontaneous decay process takes place
in a time interval of duration $\tau_\text{it}\times t$
for a quantum state with $n_q/2$ excited qubits. This estimate is based on the
assumption that 
a fidelity of unity is associated with those particular statistical realizations for which no
spontaneous decay takes place during the time interval
$\tau_\text{it}\times t$. All other statistical realizations are assumed to yield
a zero fidelity.
The estimate for $f_\text{ec1}(t)$ of Eq.(\ref{approxwith}) is based on the analogous assumption that
a unit fidelity is associated with those statistical realizations only in which no spontaneous
decay process
takes place during a recovery operation requiring a time $\tau_\text{rec}$.
If only one spontaneous decay process were possible,
the associated probability would be given by
$p=\exp (-  {\kappa}  \tau_\text{rec} n_q/2 )$ for $n_q/2$ excited qubits.
However, on the average there are $ N = {\kappa} \tau_\text{it} t n_q/2$ 
spontaneous decay processes taking place during the time interval of interest. Provided these decays are statistically independent
the mean fidelity can be estimated by $p^N$ which yields Eq.(\ref{approxwith}).
From the approximate relations \eqref{approxwithout} and \eqref{approxwith} it is apparent that the use of the one-error correcting
code space of Eq.(\ref{KL}) together with the quantum gates of Figs.\ref{fig::sqgates} and \ref{fig::tqgates}
implies a significant increase of the fidelity as long as
the average time between two successive spontaneous decay processes 
is much larger than the recovery time $\tau_\text{rec} = n_q \tau_\text{cnot} - \pi/2$.
For the duration $\tau_\text{cnot}$ of the CNOT-operations involved in
a recovery operation this condition yields 
\begin{equation}
\tau_{cnot} \ll \frac{1}{ 4\kappa}\frac{n_{\!L}}{(n_{\!L} + 1)^3}.
\label{cnot}
\end{equation} 
Thus,
for large numbers of logical qubits $n_{\!L}$ it may be difficult to realize such fast CNOT-operations.

Relation (\ref{approxwith}) is applicable if spontaneous decay processes
occurring during recovery operations are the main source of decoherence.
%
%
Using an encoding based on groups of physical qubits, i.e. quantum registers, each of which involves
its own one-error correcting code space at least some of the
spontaneous decay processes occurring during  recovery operations can be corrected provided they take place in different registers.
This way an additional significant improvement of the stabilization of quantum algorithms may be obtained if spontaneous decay
during recovery operations is the main reason for decoherence.
For this purpose
let us consider $N_\text{reg}$
basic quantum registers
each of which consists of $n_\text{reg}^{(i)}$ physical qubits and 
encodes $n_{\!L}^{(i)} = (n_\text{reg}^{(i)} -2)/2$ logical qubits
according to Eq.(\ref{KL}).
Local operations within each quantum register can be performed with the help of the universal quantum gates of
Figs. \ref{fig::sqgates} and \ref{fig::tqgates}. Entanglement operations between different quantum registers
can be performed with the entanglement gate of Fig.\ref{fig::cpgates}.
In such an architecture of a quantum information processor
all spontaneous decay processes occurring in different quantum registers can be corrected irrespective of 
whether they occur successively or simultaneously.
According to the reasoning used for the derivation of Eq. (\ref{approxwith}) the mean fidelity decay of such an encoding
can be estimated by
\begin{eqnarray}
f(t) &=& \prod_{i=1}^{N_\text{reg}} (e^{-(\kappa/2)n_\text{reg}^{(i)} \tau^{(i)}_\text{rec}})^{N^{(i)}}\equiv
e^{-R_\text{ec2} \tau_\text{it} t}
\nonumber\\
\label{block}
\end{eqnarray}
with the fidelity decay rate
\begin{eqnarray}
R_\text{ec2}  &=&
(\kappa/2)^2
\sum_{i=1}^{N_\text{reg}}\tau^{(i)}_\text{rec}(n_\text{reg}^{(i)})^2, 
\label{rate}
\end{eqnarray}
the recovery time $\tau_\text{rec}^{(i)} = n_\text{reg}^{(i)} \tau_\text{cnot} - \pi/2$ and
with $N^{(i)} = (\kappa/2) n_\text{reg}^{(i)}\tau_\text{it} t$
denoting the mean number of spontaneous decay processes taking place within register $i$
during $t$ iterations of the quantum map. 
If all quantum registers involve the same total number of logical qubits $n_{\!L}^{(i)}$,
according to Eqs.(\ref{approxwithout}) and (\ref{block})
a blockwise encoding implies a significant increase of fidelity as long as the duration of the
CNOT-operations involved in the recovery operations are small enough, i.e.
\begin{eqnarray}
\tau_\text{cnot} \ll \frac{1}{ 4\kappa}\frac{n_{\!L}^{(i)}}{(n_{\!L}^{(i)} + 1)^3}.
\label{ratio}
\end{eqnarray}
Thus, contrary to the rather stringent condition (\ref{cnot})
for $n_{\!L}^{(i)} =1$, for example, successful error correction by blocks of one-error correcting jump codes
requires only that the duration of a CNOT-operation $\tau_\text{cnot}$
is much smaller than the mean life time of a single qubit $1/\kappa$. 
In realistic applications it should not be too difficult to fulfil this condition.
So, despite its higher redundancy for any fixed total number of logical qubits $n_{\!L}$
a blockwise encoding based on $n_{\!L}/n_{\!L}^{(i)}$ quantum registers
each of which contains $n_{\!L}^{(i)}\ll n_{\!L}$ logical qubits is more stable against spontaneous decay processes than a direct encoding 
of these $n_{\!L}$ logical qubits with $n_q = 2n_{\!L} +2$ physical qubits according to Eq.(\ref{KL}).

\section{Summary and conclusions}
One-error correction jump codes are an efficient method to stabilize quantum algorithms against spontaneous decay processes provided one is capable of correcting also errors taking place during elementary quantum gates.
With the help of appropriate subspaces of these error correcting code spaces simple Hamiltonian universal quantum gates can be constructed which achieve this goal.
A realistic simulation was presented in which the quantum algorithm of the tent-map was decomposed into these quantum gates.
Even if the recovery operations cannot be implemented ideally this encoding suppresses the decohering influence of spontaneous decay processes significantly.
This error suppression can be increased with the help of a block-encoding which involves different error correcting code spaces.
For this purpose the presented universal entanglement gate might turn out to be useful as it allows to suppress simultaneous spontaneous decay processes provided they affect physical qubits of different error-correcting logical subspaces.

\begin{acknowledgments}
This work is supported by the EU IST-FET project EDIQIP and by the DFG (SPP-QIV).
Numerical support by Stanislav Vym\v{e}tal and Igor Jex is gratefully acknowledged.
\end{acknowledgments}

\bibliography{references}

\end{document}